\documentstyle[psfig,preprint,aps,prb]{revtex}
\draft

\begin{document}

\title{Spin ladders with nonmagnetic impurities}

\author{H.-J. Mikeska, U. Neugebauer}
\address{Institut~f\"ur~Theoretische~Physik, Universit\"at~Hannover, 
         Appelstra{\ss}e 2, 30167~Hannover, Germany}

\author{U. Schollw\"ock}
\address{Sektion Physik, Ludwig-Maximilians-Universit\"at~M\"unchen, 
        Theresienstra{\ss}e 37, 80333~M\"unchen, Germany}

\maketitle

\begin{abstract} 

We investigate antiferromagnetic spin ladders with nonmagnetic
impurities by variational and numerical (Lanczos and DMRG)
methods. The interaction between the two unpaired spins opposite to
the impurities is described by an effective exchange interaction
$J_{eff}$, the magnitude of which depends on the impurity distance. The
magnitude of $J_{eff}$ is different for unpaired spins at the edges of
an open ladder and in the bulk. This difference is related to the
different distribution of the unpaired spin into the bulk of the
ladder. The numerical results are interpreted using matrix product
states. Using the DMRG we calculate the spectrum of low-lying energy
levels for up to 6 impurities and find that these spectra can be
reproduced assuming pair interactions with an accuracy of better than
10\%. We discuss the filling of the ladder gap with impurity states
and argue that in the thermodynamic limit the spin ladder with a
finite concentration of impurities always shows a Curie susceptibility
at low temperatures.

\end{abstract}


\section{Introduction} 
In the last few years spin systems consisting
of two interacting spin chains with $S=\frac{1}{2}$, now usually
called (two-legged) spin ladders have attracted considerable attention
as recently reviewed by Dagotto and Rice \cite{DagR96}. The spin
ladder with isotropic antiferromagnetic interactions exhibits a spin
gap in the excitation spectrum which has been shown to be related to
the dimer gap (af coupling on the rungs only) as well as to the
Haldane gap (strong ferromagnetic coupling on the rungs). It has been
concluded \cite{Whi96} that the isotropic spin ladder is in
the same phase as the Haldane chain - this can formally be described
using matrix product ground states in a generalized spin ladder with 
additional diagonal bonds \cite{BreMN96}.  

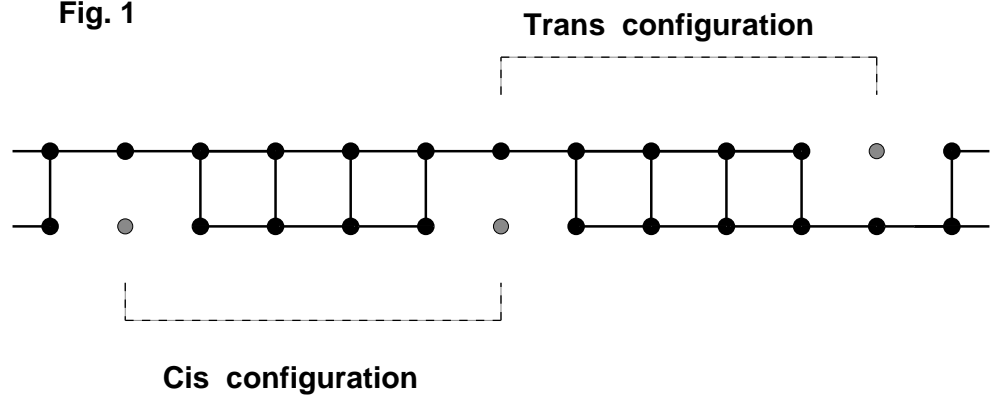
\begin{figure}
\caption{Structure of a spin ladder with three impurities leading to 
         unpaired spins in cis resp.\ trans configuration.}       
        \label{structure}
\end{figure}

In the Haldane chain, the investigation of magnetic as well as
nonmagnetic impurities has contributed substantially to our
understanding of the system: There exist quasifree spins
$S=\frac{1}{2}$ at the end of the chain segments created by the
impurities as qualitatively predicted by the valence bond ground state
\cite{AffKLT88} and experimentally found in ESR experiments on NENP
doped with Cu \cite{Kat92}. We have investigated nonmagnetic
impurities in spin ladders and found that these lead to effects both
similar to and different from impurities in the Haldane chain. The
essential difference is of geometric origin: impurities in a spin
ladder do not break the sequence of magnetic interactions, but create
weak links: one spin $S=\frac{1}{2}$ on one leg of the ladder has no
counterpart on the other leg and mediates between the adjacent regular
ladder structures. In the following we use the term 'unpaired spin'
for these spins without counterpart on the opposite leg.  A typical
configuration with three impurities is shown in fig.~\ref{structure};
neighboring impurities may lead to unpaired spins on the same leg (cis
configuration) or on opposite legs (trans configuration). This
structural effect that the impurities do not break the coupling along
the linear arrangement is analogous to the one of impurities in a
chain with nn and nnn interactions - which can actually be considered
as a generalized ladder \cite{BreMN96}. For a statistical
distribution of impurities both the sequence of cis and trans pairs
and the distances between these pairs will be statistical. The
distance between neighboring unpaired spins is conveniently counted by
the number $p$ of complete rungs between them. Assuming small impurity
concentration $c$ we neglect the possibility of complete breaking of
the ladder which is proportional $c^2$.

For a microscopic understanding of the effects of a low concentration
of defects on the spectrum of the spin ladder we present in the
following numerical and analytical calculations for the properties of
the ground state and of low-lying excited states of impure ladders
with $S=\frac{1}{2}$ and all exchange interactions antiferromagnetic,
isotropic and of equal magnitude (which is set equal to unity). In
particular we have considered:

\begin{itemize}
\item
two spins $S=\frac{1}{2}$ in either cis or trans configuration at the
edges of an open ladder formed by $p$ rungs (exact diagonalization with
the Lanczos algorithm and analytical estimates),

\item
two impurities at varying distances in either cis or trans
configuration in the bulk ladder (exact diagonalization with the
Lanzcos algorithm for periodic boundary conditions, DMRG calculations
for open boundary conditions and estimates from matrix product
states),

\item
the spin ladder with many impurities (up to six impurities in DMRG
calculations and arguments based on the Lieb-Mattis theorem).

\end{itemize}

Our work is motivated by recent experiments \cite{AzuFTIONT96}
on the ladder material $\rm SrCu_2O_3$ doped with $\rm Zn$, which show
a tendency of the ladder gap to vanish with increasing $\rm Zn$
concentration. Calculations for similar systems which, however,
concentrate on somewhat different aspects have recently been done by
Motome et al \cite{MotKFI96} and Martins et al \cite{MarDR96}.

\section{Numerical Methods} 

We have performed exact diagonalizations using the Lanczos technique
for spin ladder systems with two defects and a varying number $p \le
12$ of complete rungs, i.e.\ $N=2p+2$ spins in total and have
considered both periodic and open boundary conditions. For open
boundary conditions we study a ladder with two impurities at the
edges of the system and consequently separated by $p$ rungs in either
cis or trans configuration. In the case of periodic boundary
conditions we fix one defect on the first rung and vary the distance
($p_1, \; p_1 \leq [p/2]$ rungs) to the second defect. We again
consider cis and trans configurations with $p=11$ and $p=12$
rungs. For these configurations we have calculated the energies of the
ground state and of low lying excitations and, for open boundary
conditions only, the distribution of the magnetization.

For the normal ladder system projection of the Hilbert space on the
irreducible representation of the corresponding symmetry groups
(reflection and mirror symmetry, total SU(2) invariance and, in the
case of periodic boundary conditions, translational symmetry)
drastically reduces the amount of memory for diagonalization within
these subspaces. As a conseqence of inserting impurities into the
ladder most of these symmetries, in particular translational
invariance, are no longer present and the computational effort is
correspondingly larger. It is only the SU(2) invariance of the model
which is preserved.

The calculations were performed on a MPP CRAY T3D SC256 of the Zuse 
Computing Centre Berlin using a parallel implementation of the 
Lanczos algorithm. The numerical accuracy of these calculations is
$10^{-10}$ or better.

Whereas the accuracy of the exact diagonalizations using the Lanzcos
approach is high, it can only be done for relatively small systems.
To study longer chains we have used the Density Matrix Renormalization
Group (DMRG) \cite{Whi92}. Using this method we have studied open
ladders from $2 \times 50$ up to $2 \times 100$ spins. As the defects
are found to behave like spins localized on the bulk correlation
length scale, defects more than 20 sites away from the ladder ends
behave effectively like in an infinite system.

As good quantum number we used the total $S^{z}$ spin; in
configurations with defects distributed symmetrically around the
ladder center, we also used parity. This allows for fast
classification of states.

We typically kept $M=100$ to $M=150$ block states. A remark is in
order on the precision of the DMRG in this particular application. For
the defect-free spin ladder, truncation errors are very low, for
$M=100$ $\rho=2.9 \times 10^{-11}$ and for $M=150$ $\rho=1.1 \times
10^{-12}$, corresponding to an extremely small error in the energies
of the defect-free system for low-energy states. The effective
precision of the DMRG is however greatly reduced, when a defect or
defect pair are added. The block states from the step before are not
optimally chosen to represent the altered system. We find that the
introduction of defects effectively reduces the precision of the DMRG
by up to several orders of magnitude.

This problem becomes particularly pressing in defect configurations
{\em nonsymmetric} with respect to the ladder center. For any even
number of defects, the ground state is the lowest eigen state in the
$S^{z}=0$ sector. A nonsymmetric defect configuration implies that
during the growth process there will be an odd number of defects
present in some DMRG steps. The total number of spins on the ladder
changes from even to odd. Consequently, the ground state jumps to the
$S^{z}=\pm \frac{1}{2}$ sector, to which the available block states
are not well adapted, drastically reducing the precision of the DMRG.
However, the error can be greatly reduced by using the finite size
DMRG algorithm \cite{Whi93,SchJ95}: The whole chain is recalculated in the
presence of the complete defect configuration without any jumps
between different $S^{z}$ sectors. The gain in precision by far
exceeds the one obtained by increasing $M$.

To illustrate this point, let us consider the case of a $2 \times 50$
spin ladder with two defects in trans-configuration on rungs 24 and
26. We have calculated the lowest eigenstates with $S^{z}=0$ and
$S^{z}=1$, keeping $M=40$ and $M=100$ states. In Table~\ref{tabdmrg}
the energies given are for the chain calculated by conventional DMRG,
and after one resp.\ two iterations applying the finite size
algorithm. Obviously, even for $M=100$ the unmodified DMRG (first
line) gives results of the order of 0.01 away from the converged
result, whereas the finite size algorithm produces highly precise
results even for $M=40$ (compare results in second and third lines).

All DMRG calculations were performed on a PentiumPro 200MHz machine running
under Linux.

\section{Two unpaired spins} 

In this section we discuss the spectra and spin configurations for two
unpaired spins in ladders with both open and periodic boundary
conditions, presenting and comparing numerical results from both Lanczos
and DMRG calculations and from analytical approaches. 

\subsection*{Numerical results for the spectrum of unpaired edge spins}

We consider the configuration with $p$ rungs connecting two spins
$\vec S, \vec S'$ in either cis or trans configuration at the
ends. Using the Lanczos algorithm we have exactly diagonalized systems
with $p \le 12$. We find that there is an energy gap of the order of
the known ladder gap ($\Delta \approx 0.53$) but the ground state is
splitted into a singlet and a triplet state. As a consequence of the
Lieb-Mattis theorem the lowest state for our configuration is a
singlet if the two unpaired spins are on different sublattices ($p$
even for cis spins and $p$ odd for trans spins) and a triplet if they
are on the same sublattice ($p$ odd for cis spins and $p$ even for
trans spins) as discussed in ref. \onlinecite{SigF96}. We interpret
this splitting as resulting from an effective interaction between the
unpaired spins $\vec S, \vec S'$ at the ladder boundaries and we
derive an effective coupling by writing a Hamiltonian in the subspace
of these lowest two states as

\begin{equation} H_{\pm}^{(p)} = E_{0,\pm} \ + J_{eff,\pm}^{edge}(p)
\: \vec S \: \vec S'.  
\end{equation} 

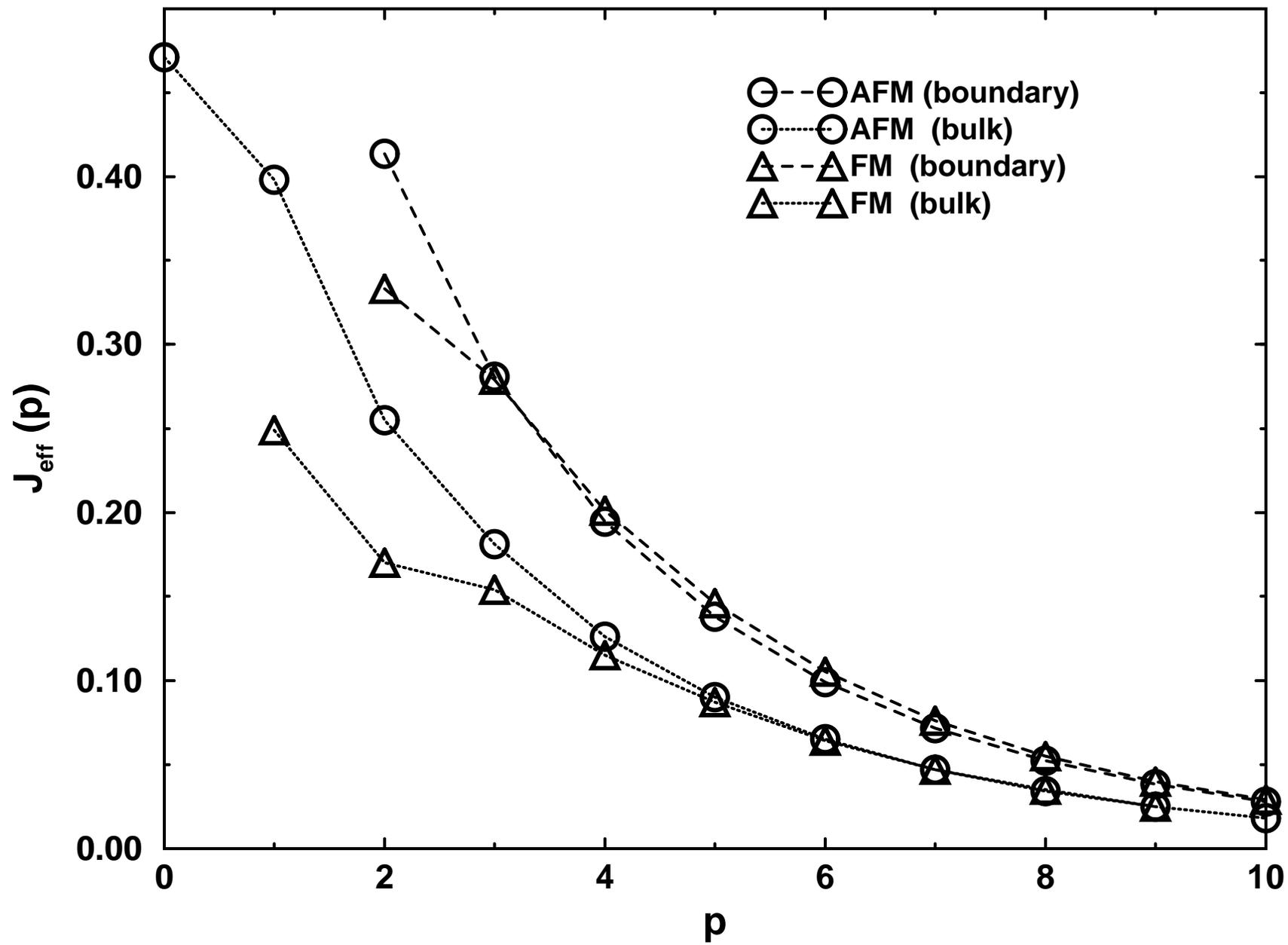
\begin{figure}
\caption{Effective exchange interactions for two unpaired spins at the 
         edges and in the bulk of a spin ladder. Exchange energies for 
         effective ferromagnetic and antiferromagnetic interactions are 
         plotted separately.}
         \label{resultopen}
\end{figure}

The index $\pm$ refers to the sign of $J_{eff}$, i.e.\
antiferromagnetic and ferromagnetic effective interaction
respectively. $J_{eff}^{edge,\pm}(p)$ is given in
Table~\ref{tabJopen} and is also plotted in fig.~\ref{resultopen} to
show its dependence on $p$. $J_{eff,\pm}^{edge}$ actually is identical
to the singlet-triplet splitting of the ground state. An excellent fit
for $p \ge 5$ is obtained as

\begin{eqnarray} 
E_{0,+}(p) &=& -1.157 \ p, \nonumber \\ 
J_{eff,\pm}^{edge}(p)
&=& J_{0,\pm} \ e^{-p/\xi}, \qquad {\rm with} \qquad J_{0,+} \approx
0.674, \quad J_{0,-} \approx 0.714, \quad \xi \approx 3.1.
\end{eqnarray} 
The uncertainty in these data as determined from a least square fit is
$10^{-6}$ for $J_{0,\pm}$ and better for $E_{0,+}$. Due to the finite
size of the open chain, $E_{0,-}$ ($\approx -1.161 \ p$) cannot be
determined with equal accuracy. We draw attention to the fact that the
best fit is purely exponentially decaying and that a behavior 
$\propto exp(-p/\xi)/\sqrt p$ can be excluded.  This is in parallel to the
behavior of an open Haldane chain \cite{Whi92}.

\subsection*{Spectrum of unpaired spins in the bulk ladder: Numerical results}

Contrary to the Lanczos approach the DMRG allows to deal with
sufficiently long ladders so that the two unpaired spins can be at some
rather large distance and at the same time sufficiently far from the
boundaries to identify their bulk interaction. The splitting of the
ground state for two unpaired spins is again described by a
Hamiltonian as above with $J_{eff,\pm}^{edge}(p)$ replaced by
$J_{eff,\pm}^{bulk}(p)$. The latter quantity is also given in
Table~\ref{tabJopen} and shown in fig.~\ref{resultopen}.
We find (for $p \ge 5$)

\begin{equation}
J_{eff,\pm}^{bulk}(p) \ = \ J_{0,\pm}^{bulk} \ e^{-p/\xi}, \qquad {\rm with} 
\qquad J_{0,+}^{bulk} \ \approx \ 0.43, \qquad \xi \approx 3.1.
\end{equation}
$J_{0,-}^{bulk}$ is approximately equal to $J_{0,+}^{bulk}$, but
determined less accurately. We notice that $J_{eff}^{bulk}(p)$ is
reduced with respect to $J_{eff}^{edge}(p)$ by a reduction factor
$r_{exc,\pm}$,

\begin{equation} 
J_{eff,\pm}^{bulk}(p) \ = r_{exc,\pm}(p) \ J_{eff,\pm}^{edge}(p).
\end{equation} 

For effectively antiferromagnetic interaction the reduction factor
$r_{exc}$ is $p$ independent, $r_{exc} \approx 0.65$, for effectively
ferromagnetic interaction it approaches this value asymptotically.
This reduction will be explained quantitatively below; qualitatively
it is due to the fact that the unpaired spin projection delocalizes
both into its right and left neighborhood in the bulk case whereas
there are neighbors on one side only in the boundary case. The
dependence of the effective exchange on distance is again purely
exponential and the prediction in ref. \onlinecite{SigF96}, which
includes an additional factor $1 / \sqrt p$ is not verified from our
data. The 'best' fit enforcing the factor $1 / \sqrt p$ ends up with
$\xi \approx 4$, i.e.\ a correlation length which is not appropriate.

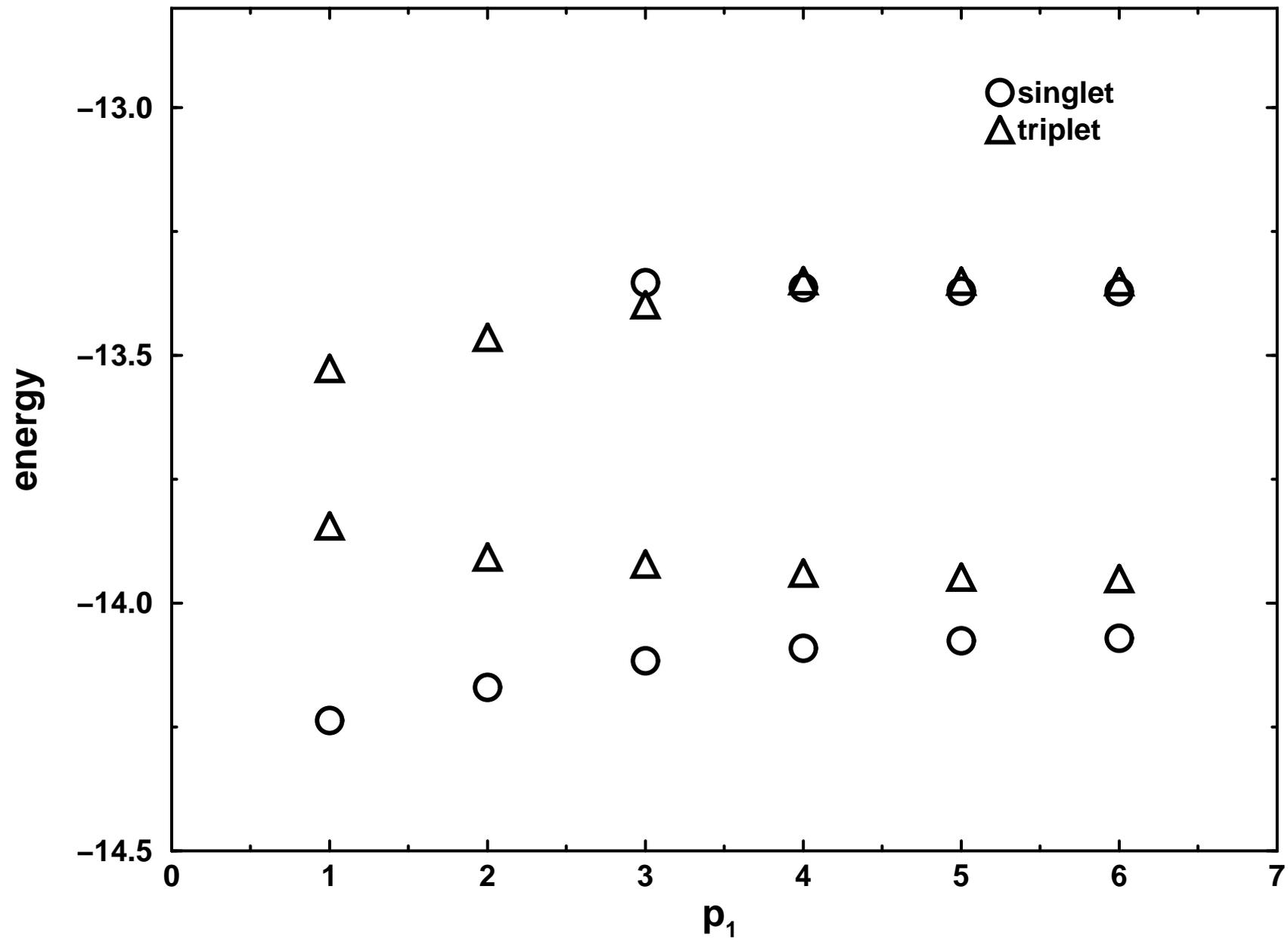
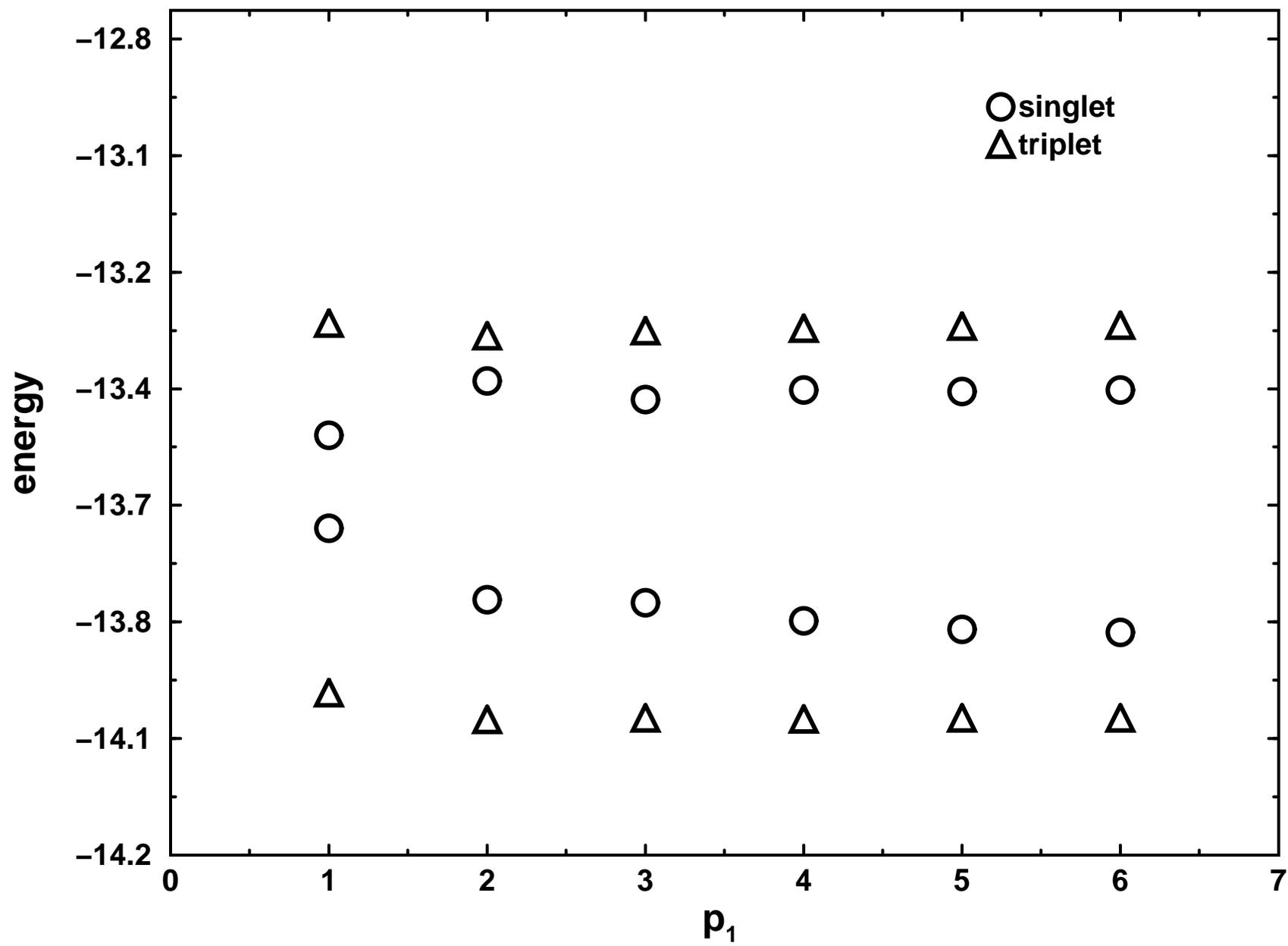
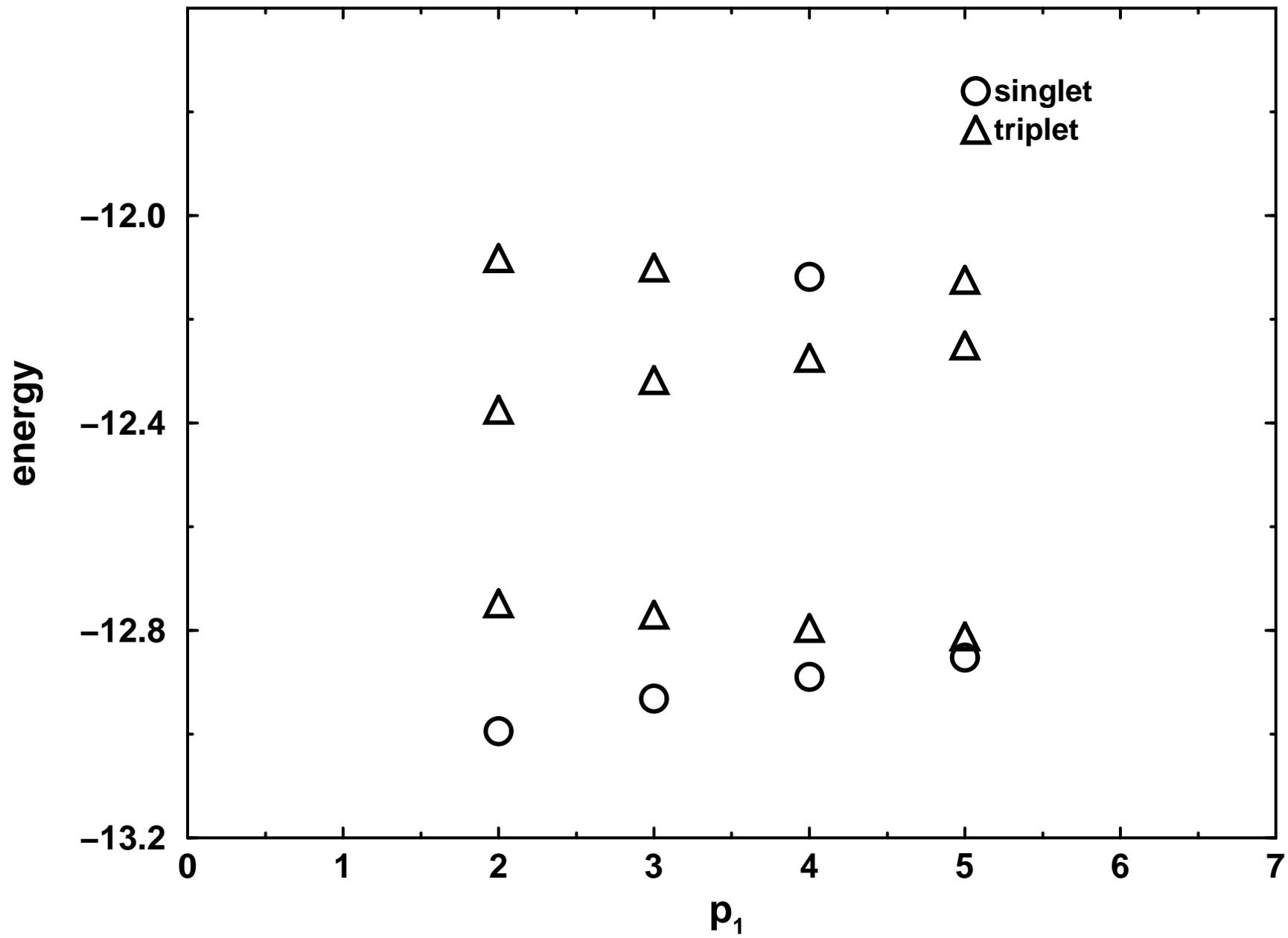
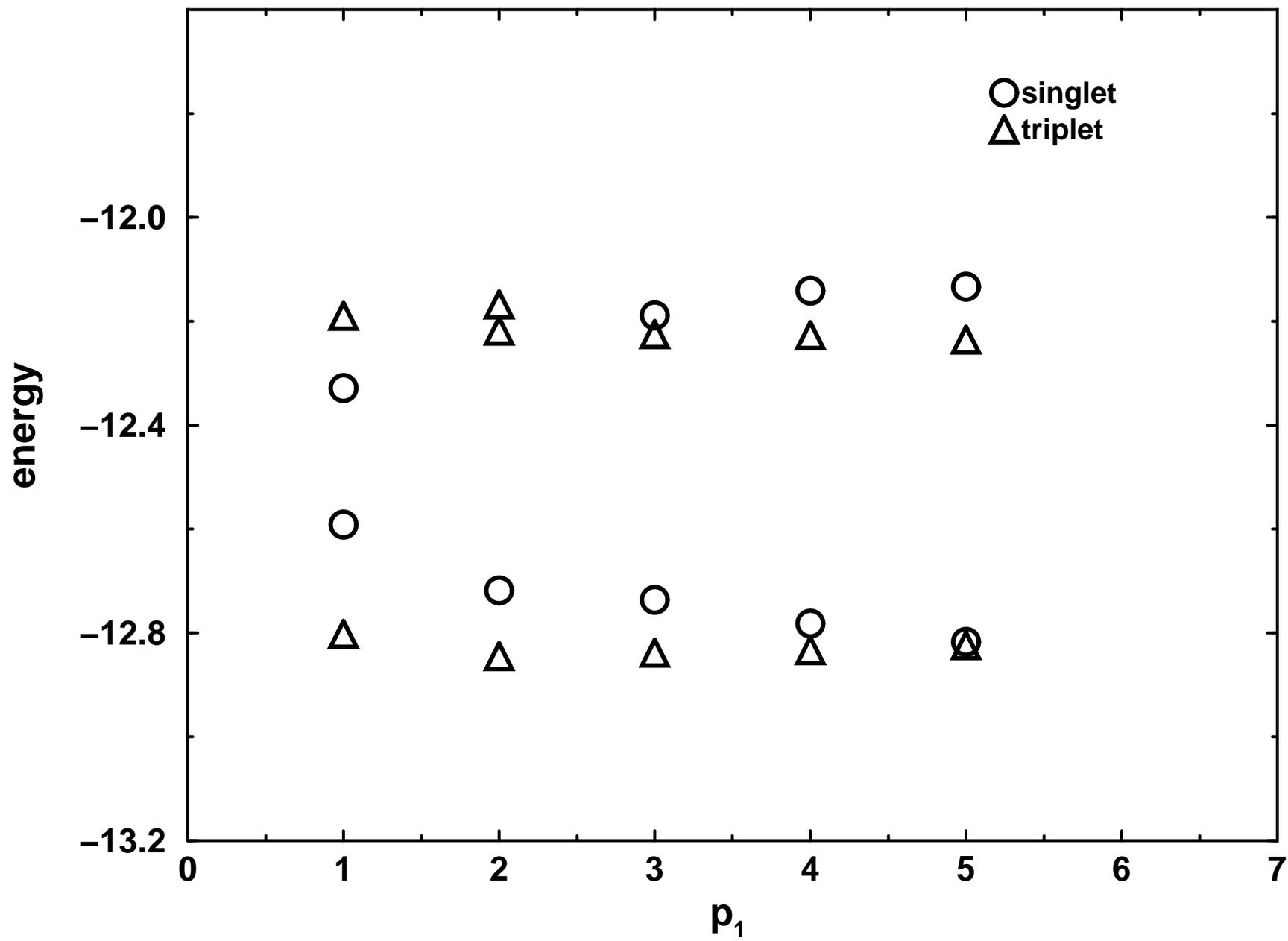
\begin{figure}
\caption{Low-lying energy levels for ladders with periodic boundary 
        conditions and two unpaired spins with a distance of $p_1$ resp.\
        $p_2$ complete rungs. Spectra for unpaired spins with 
        effective antiferromagnetic, resp.\ ferromagnetic, interactions
        are shown separately. (a, b) $p_1+p_2 = 12$ rungs, 
        (c, d) $p_1+p_2 = 11$.}
        \label{spectrumpbc}
\end{figure}

Using the Lanczos algorithm, we have also done exact diagonalizations
for configurations with periodic boundary conditions (pbc) and two
spins $\vec S, \vec S'$ in cis or trans configuration separated by
$p_1$, resp.\ $p_2$ rungs. The spectra are shown in
fig.~\ref{spectrumpbc}a,b for $p_1 + p_2 = 12$ and in
fig.~\ref{spectrumpbc}c,d for $p_1 + p_2 = 11$; their qualitative
structure is the same as for the open ladder. We find that the smaller
of the two values $p_1, p_2$ determines whether the ground state is
singlet or triplet when the rules given above for the open ladder are
used. From the graphs presented in fig.~\ref{spectrumpbc} it is also
evident that a much larger singlet-triplet splitting of the ground
state is obtained for $p_1 + p_2$ even than for $p_1 + p_2$ odd. The
obvious reason is that the two exchange interactions which have to be
added for pbc have equal (different) sign for $p_1 + p_2$ even
(odd). Using again the concept of an effective exchange interaction,
$J_{eff}^{pbc}(p_1,p_2)$, we expect for two unpaired spins (resulting
from two impurities in a periodic ladder)

\begin{equation}      
J_{eff}^{pbc}(p_1,p_2) = J_{eff}^{bulk}(p_1) + J_{eff}^{bulk}(p_2).
\end{equation} 
Inserting the results as given before, we find that this relation is
obeyed to within 10\%.

In fig.~\ref{spectrumpbc} we also show the results for the lowest
excited states, i.e.\ in particular the splitting of the regular
ladder gap with energy $\approx \Delta$. For most of the defect
configurations the effective interactions have different sign for the
ground state and for the first excited state (energy of the regular
ladder gap). For example, a singlet ground state is related to a
triplet as the lower one of the states of energy $\approx \Delta$, as
follows from a simple coupling of the impurities to the lowest bulk
excitation with $S=1$. This rule, however, is not strictly obeyed and
a detailed discussion of the structure of the excited states in the
impure ladder has to be reserved for future work.

\subsection*{Analytical approaches to unpaired spins on the ladder edges}

A qualitative understanding of our numerical results for the open
ladder can be obtained using matrix product wave functions as
described in ref.\onlinecite{BreMN96}. We consider a system with two
spins at the ladder edges in trans configuration and $p$ rungs in
between; A wave function describing this system with four degrees of
freedom can be written down as a $2 \times 2$ matrix product (MP) wave
function,
 
\begin{equation}
        |\psi \rangle _{\sigma, \sigma'} = 
        \left[ \prod_{j=1}^{p+1} g_j \right]_{\sigma, \sigma'},
        \qquad g_j = \left( \begin{array}{cc}
        a|t_0 \rangle_j + b|s \rangle_j & -a \sqrt{2} |t_+ \rangle_j \\
        a \sqrt{2} |t_- \rangle_j & -(a|t_0 \rangle_j- b|s \rangle_j)
        \end{array} \right). \label{mptrans} 
\end{equation} 
Here the matrix $g_j$ describes the coupling of spins situated on
diagonal sites on two adjacent rungs to singlets, resp.\ triplets; the
boundary spins are coupled likewise to the neighboring rung. The
value of $b^2 = 1- 3 a^2$ is determined by minimization of the energy;
for the numerical estimates below we take the result for the infinite
ladder \cite{BreMN96}, $b \approx 0.1735$. We have chosen the two edge
spins in trans configuration where a MP wave function is easily
written down, whereas this is more difficult for two spins in cis
configuration (with e.g. two more spins on the upper leg than on the
lower leg); the latter case requires a more detailed analysis which
will be published separately. Instructive limiting cases of the
wavefunction of eq.(\ref{mptrans}) are:

\begin{itemize}

\item For $a=b=\frac{1}{2}$ it is the wave function for $p$ singlets
on the $p$ rungs and truly free boundary spins $\vec S, \vec S'$. This
wave function actually is an eigenfunction to the Majumdar-Ghosh
Hamiltonian for a finite chain with $2p$ spins and one additional free
spin at each open end. Their spin projections can be identified 
with the matrix indices $\sigma, \sigma'$.

\item For $b=0, a=1/\sqrt3, \quad |\psi \rangle _{\sigma, \sigma'}$
describes $p+1$ units with $S=1$ on the diagonals of the ladder, i.e.\
it is identical to the 4 eigenfunctions for the AKLT chain with open
ends \cite{AffKLT88}. In this limit it is known that at the edges
there exist quasifree $S=\frac{1}{2}$ spins which extend somewhat into
the bulk of the ladder. Note that the matrix indices $\sigma, \sigma'$
now can no more be identified with the spin projections of the edge 
spins. 

\item The general case will be intermediate and the boundary spins extend
into the adjacent bulk ladder to an extent determined by the correlation
length. This correlation length is underestimated by the MP ansatz but the
tendency of the variation is given correctly: it decreases with
increasing $b$, i.e. from the AKLT limit via the ladder configuration to
the Majumdar Ghosh limit. 

\end{itemize}

For a more quantitative treatment we couple the 4 MP wavefunctions
$\left[ \prod_j g \right]_{\sigma,\sigma'}$, $\sigma, \sigma' = \pm
1$, to singlet, resp.\ triplet states and calculate the effective
exchange energy from the energy difference of these states. This leads
to the following results for a configuration with $p$ complete rungs
in between:

\begin{equation} 
J_{eff,MP}^{edge}(p) = \pm J_0 \ e^{-p/\xi}, \qquad
{\rm with} \qquad J_0 \approx 0.77, \quad \xi^{-1} \approx 1.23.
\end{equation} 
This gives the correct sign of $J_{eff}$
(i.e.\ ferro- or antiferromagnetic if $\sigma,\sigma'$ are on the same
or on different sublattices respectively) and a remarkably good
numerical value for $p=0$; the decay with distance, however, is too
strong, due to the fact that the correlation length is underestimated
in MP states.

We notice that the low energy structure of the spectrum is the
generalization to spin ladders of the Kennedy triplet observed in
Haldane chains with open boundary conditions \cite{Ken90}: The latter
is obtained (in the AKLT limit) by taking $b \to 0$ in our
wavefunction. In the ladder the wavefunction is more general than just
a RVB ansatz, but the effect of the quasifree boundary spins is the
same. In this sense our results are related to the results of Hida
\cite{Hid95} who investigated the ladder with ferromagnetic coupling
$\lambda$ on the rungs. This coupling connects two spins which in our
antiferromagnetic ladder are not coupled directly but as nnn by two af
bonds; from our calculations we see that this leads to the same
effects.

An alternative approach to arrive at a theoretical estimate for
$J_{eff,\pm}^{edge}$ is to integrate out the spin degrees of freedom on
the rungs between the boundary spins. For this purpose we write the
general Hamiltonian with boundary spins $\vec S_0$ and $\vec S_{2p+1}$
as

\begin{equation}
H = H_{ladder}(1,2...2p) + h_{0,1} + h_{2p,2p+1}. 
\end{equation}
We eliminate $h_{0,1}$ and $h_{2p,2p+1}$ to first order by a suitable
canonical transformation and average over the eigenstates of the
complete rungs forming the intermediate part of the ladder to end up
with an effective low-energy Hamiltonian of the form

\begin{equation}
H_{\pm}^{(p)} = E_{0,\pm}(p) \: 
                    + \: J_{eff,\pm}(p) \: \vec S_0 \: \vec S_{2p+1}.
\end{equation}
The explicit results for two and three intermediate rungs between the 
boundary spins are

\begin{eqnarray}
E_{0,\pm}(2) &=& - \frac{5}{16}, \qquad J_{eff,+}^{edge}(2) = \frac{1}{3}, 
\qquad J_{eff,-}^{edge}(2) = -\frac{1}{4}, \nonumber \\
E_{0,\pm}(3) &=& - 0.344, \qquad J_{eff,+}^{edge}(3) = - 0.2377,  \qquad 
J_{eff,-}^{edge}(3) = 0.2248.
\end{eqnarray} 

These results agree to the numerical data within 20\%. For two
isolated spins, higher order contributions to the canonical
transformation will change the numbers towards the correct values but
due to isotropy and to the fact that we are dealing with
$S=\frac{1}{2}$ the form of the Hamiltonian will not be
affected. Although this approach gives only rough estimates, it serves
to illustrate that for more than two unpaired spins higher order terms
in the canonical transformation will lead to pair interactions between
unpaired spins which are not nearest neighbors and to $m$-spin
interactions ($m \ge 4$). The difference between the above analytical
result and the numerical results should be considered as an indication
that these higher spin interactions cannot be neglected at the outset
and a description of the impure chain in terms of an effective pair
Hamiltonian for unpaired spins as in ref.~\onlinecite{SigF96} needs
justification. We will return to this question in section IV.

\subsection*{Spin configurations in the presence of two unpaired spins}

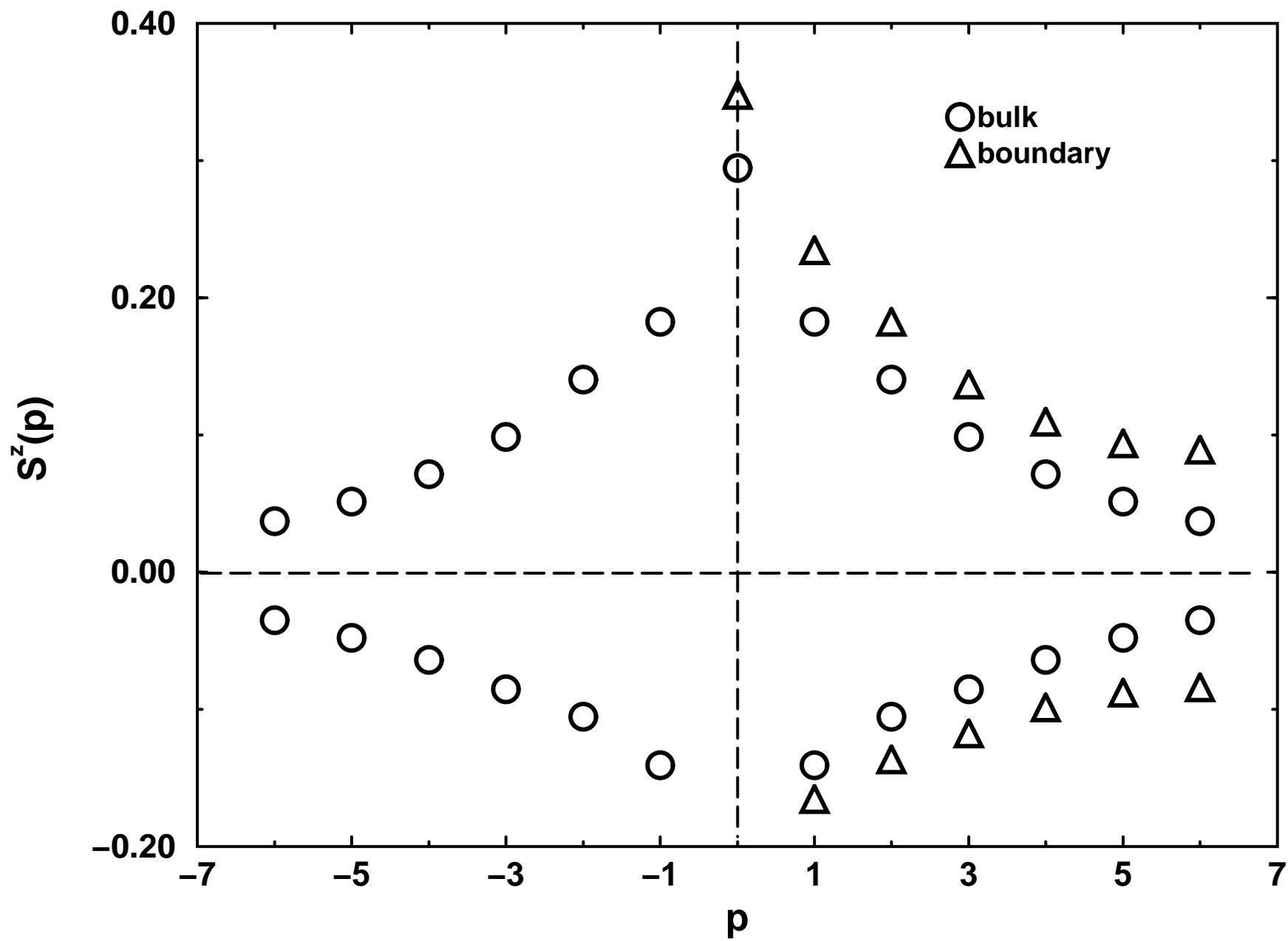
\begin{figure}
\caption{Distribution of the excess z-component of spin (magnitude 1/2) 
        resulting from the unpaired spin opposite to an impurity into
        the adjacent rungs of the ladder (the positive and negative 
        contributions for each rung $p$ are alternating between the two
        legs).}  
        \label{spinspread} 
\end{figure}

We now discuss in more detail the spin configuration in the presence
of two unpaired spins separated by a number of rungs larger than the
correlation length. Then the spin projection will spread into the
adjacent part of the ladder, i.e.\ into the rungs to the right and to
the left of its site (or into the rungs on one side only in the case
of edge spins). In order to demonstrate this effect for an unpaired
spin with a given projection $\langle S^z \rangle = \pm {1 \over 2}$
we consider the quantity $\langle S^z_{m,\pm} \rangle$, where $m$ is
the distance from the unpaired spin ($m=0$ denotes the unpaired spin)
and $\pm$ distinguishes between spins on the same (+), resp.\ opposite
(-), sublattice as the unpaired spin. When the two unpaired spins are
sufficiently separated we obtain a clear picture of the situation in
the neighborhood of a single unpaired spin coupling the two unpaired
spins to $S_{tot}=1$ and considering states with $S^z_{tot} =
+1$. Results from the DMRG and Lanczos which illustrate this
redistribuiton of the spin projection are shown in
fig.~\ref{spinspread}; all data are of the following form:

\begin{eqnarray*}
\langle S_{m,\alpha}^z \rangle \ &=& \ s_{center}^z 
                                 \quad {\rm for} \quad m=0  \\    
        &=& \ s_{tail}^z \ e^{- {\vert m \vert \over \xi}} 
        \qquad {\rm for} \qquad \vert m \vert > 0, \: \alpha = + \\
        &=& \ - \ s_{tail}^z \ e^{- {\vert m \vert \over \xi}} \ + \ \delta_m
        \qquad {\rm for} \qquad \vert m \vert > 0, \: \alpha = - \\
\end{eqnarray*}
When the two unpaired spins are sufficiently far apart, the total spin 
projection $+ {1 \over 2}$ has to be recovered, i.e.\ 

\begin{eqnarray*}
s_{center}^z + 2 \ \sum_{m=1}^{\infty}\delta_p  = {1 \over 2} 
\end{eqnarray*}
for an unpaired spin in the bulk, and
 
\begin{equation}
s_{center}^z + \ \sum_{m=1}^{\infty}\delta_p = {1 \over 2} 
\end{equation}
for an unpaired spin at the edge. $\delta_p$ is a correction which
approaches zero rapidly for $p > 5$. Quantitative results from the
two approaches are:

For an unpaired spin in the bulk (from DMRG):

\begin{equation}
\xi = 3.1 \qquad  s_{center}^z = 0.294 \qquad s^z_{tail} = 0.247,
\end{equation}
and for an unpaired spin at the edge (from Lanczos, after corrections
for the finite size of the system, i.e.\ the contributions of the second
impurity):

\begin{equation}
\xi = 3.1 \qquad s_{center}^z = 0.348 \qquad s^z_{tail} = 0.308.
\end{equation}

The amplitude $s^z_{tail}$ characterizes the redistribution of spin
projection into the adjacent rungs and is seen to be rather large.
The total excess spin on the positive sublattice is 0.894 (bulk),
resp.\ 1.117 (edge) and on the negative sublattice -0.792 (bulk),
resp.\ -0.965 (edge). We notice that the asymptotic exponential
behavior governed by the correlation length is correct for all sites
on the sublattice of the unpaired spin whereas it is approached within 
typically 5 rungs on the opposite sublattice. 

The numbers given above are consistent with the difference between the
effective exchange constants $J^{edge}$ and $J^{bulk}$ noticed above
and give a microscopic understanding of this difference as is seen in
the following way: For an unpaired spin in the bulk of the ladder the
magnitudes of both $s^z_{center}$ and $s^z_{tail}$ are reduced by a
factor of $\approx .8$ as compared to an unpaired spin at the
ladder edge. Since the effective interaction of two unpaired spins
will be determined by the tails of the spin distribution, $J_{eff}$
will be smaller for unpaired spins in the bulk ladder by 

\begin{equation}
\left( \frac{s^{z,bulk}_{tail}}{s^{z,edge}_{tail}} \right) ^2 \  \approx 0.64, 
\end{equation}
consistent with the value $r_{exc} \approx .65$ found from the analysis 
of the energy levels above.

A redistribution of $S^z$ in the neighborhood of an unpaired edge
spin is also obtained from the MP wave function above which gives
  
\begin{equation}
s^z_{center} \ = \ a^2 \ + \ ab \ \approx \ 0.422.
\end{equation}

The behavior on the adjacent rungs is different in detail from the
numerical results: $S^z_p$ decays with alternating sign and purely
exponential with different amplitudes on the leg of the unpaired
edge spin ($\tilde s^z_{tail} = a^2 + ab \approx 0.422$) and on the 
opposite leg ($\tilde s^z_{tail} = -a/(a+b) \approx - 0.767$).
Again, the MP approach gives a reasonable qualitative picture but
fails quantitatively.

\section{Many Impurities}

It is natural to assume that the results obtained for two unpaired
spins can be generalized to give results for the low energy properties
of the general impure ladder with impurity concentration $c$ by using
pair exchange interactions between unpaired spins which are nearest
neighbors only and determining sign and magnitude of these
interactions from the results in section III. For a random
distribution of impurities this means that the impure ladder reduces
to a spin chain with exchange constants which are random with respect
to both sign and magnitude. This is the model used in
ref.~\onlinecite{SigF96}.

In this section we will investigate this assumption more closely by
comparing exact (DMRG) spectra for various configurations with four
and six impurities to spectra obtained from the assumption above. We
have restricted ourselves to an even number of impurities in order
to facilitate the comparison between different impurity concentration
(for an odd number of impurities the spin of the ground state will be
half integer). A similar approach to the case of the spin-Peierls
substance $\rm CuGeO_3$ with impurities has been done in
ref.~\onlinecite{MarDR96}.

Before presenting these numerical results we want to discuss some
aspects of the general impure ladder with antiferromagnetic exchange
interactions only and $L$ rungs, i.e.\ $2L$ sites, of which $N$ sites
are occupied by nonmagnetic impurities. The ground state of the
remaining $2L - N$ spins $S=\frac{1}{2}$ can take values $S_{tot}$ of
total spin between $0$ and $N/2$, depending on the distribution of the
impurities on the two sublattices. For a random distribution of the
impurities the distribution of $S_{tot}$ is easily obtained from the
theorem of Lieb and Mattis which states $S_{tot} = \frac{1}{2} \vert
N_A - N_B \vert$, where $N_A, N_B$ are the numbers of defects on the
$A$ and $B$ sublattices respectively (for configurations with two
defects on the same rung, which break the ladder, this gives $S_{tot}$
only for one out of a number of degenerate ground states). Thus for a 
random distribution of defects the probability $g(S_{tot})$ 
for a ground state to have total spin $S_{tot}$ is 

\begin{eqnarray}
{\rm for} \quad S_{tot} = 0 \hskip 2.0cm g(S_{tot}) &=& {N \choose N/2} 
                                                  \nonumber \\
{\rm for} \quad S_{tot} > 0 \hskip 2.0cm 
                              g(S_{tot}) &=& 2 \, {N \choose N/2 - S_{tot}}.
                \label{spinweight}
\end{eqnarray}
For $N \gg 1$ we obtain from Stirlings formula the relative weight of states
with $S_{tot}=0$ and $S_{tot}=1$ ($2^N$ is the total number of defect induced
states) 

\begin{equation}
\frac{g(S_{tot}=0)}{2^N} = \sqrt{\frac{2}{\pi N}}, \qquad 
\frac{g(S_{tot}=1)}{2^N} = 2 \ \sqrt{\frac{2}{\pi N}}. 
\label{S0value}
\end{equation}
The ground state spin value with largest probability therefore is
$S=1$. Of particular interest are the average values of the ground
state spin for a random distribution of $N \gg 1$ impurities. They are
calculated using eq.~(\ref{spinweight}) to give:

\begin{eqnarray}
\langle S_{tot} \rangle &=& \sqrt{\frac{N}{2 \pi}} \nonumber \\
\langle S_{tot}^2 \rangle &=& \frac{N}{4}
\end{eqnarray}
Thus the result obtained in ref.\ \onlinecite{SigF96} from a random
walk argument is seen to be an exact consequence of the Lieb-Mattis
theorem. For a discussion of the low temperature susceptibility in
some given defect configuration we use $S_0$ to denote the total spin
of the ground state (with zero energy) and characterize the remaining
states $\alpha = 1,2 ... \alpha_m=2^N-1$ by their total spin
$S_{\alpha}$ and excitation energy $\Delta_{\alpha} > 0$. The
limiting susceptibility for low magnetic fields is then given by

\begin{equation}
\chi = \frac{(g \mu_B)^2}{3 k_B T} \: 
 \frac{z(S_0) + \sum_{\alpha=1}^{\alpha_m} z(S_{\alpha}) 
                                             e^{- \beta \Delta_{\alpha}}}
    {(2S_0+1) + \sum_{\alpha=1}^{\alpha_m} (2 S_{\alpha} + 1) 
                                             e^{- \beta \Delta_{\alpha}}},
    \nonumber 
\end{equation}
with 

\begin{equation} z(S_{\alpha}) = 2 \ S_{\alpha} \ (S_{\alpha} +
\frac{1}{2})\ (S_{\alpha} + 1).  \end{equation} If the ground state of
the impure ladder is not a singlet, i.e.\ $S_0 > 0$, leading to
$z(S_0) > 0$, a Curie susceptibility results in the low temperature
limit. For $S_0 = 0$, leading to $z(S_0) = 0$, the susceptibility
shows an activated temperature dependence characteristic of a gapped
system ($\Delta_1$ finite); if the low-lying states get dense with an
asymptotic density of states $\rho(\epsilon) \sim \epsilon^{-\alpha}$
the low temperature behavior of the susceptibility changes to
$\chi(T) \sim T^{-\alpha}$. In particular a temperature independent
susceptibility results for $\alpha = 0$, i.e.\ a constant density of
states (this is the case of the $S=\frac{1}{2}$ Heisenberg
antiferromagnetic chain - actually it should be realized in this
context that a concentration of $c=\frac{1}{4}$ is sufficient to turn
the ladder geometry into that of a single chain). All these deviations
from a Curie behavior require a singlet ground state, which according
to eq.(\ref{S0value}) occurs with negligible weight in the limit of a
macrosocpic system. We therefore conclude that the low temperature
susceptibility of the impure ladder always follows a Curie law. This
confirms from a different point of view the result $\alpha = 1$ which
has been obtained from a renormalization group approach
\cite{WesFSL95}.

When the limiting value of the Curie constant for $T \to 0 $ is
calculated on the basis of the above expressions, we find that due to
the random positions of the defects the factor $\frac{3}{4}$
(resulting from $S (S+1)$ for $S=\frac{1}{2}$) is replaced by
$\frac{1}{4}$. The experimentally interesting behavior at finite
temperatures involves the transition between these two limiting cases,
i.e.\ a change in Curie constant by a factor of $3$. It will be
determined to a large extent by the density of states. This is not
easily accessible, and in the remainder of this section we present
what can be learned from numerical calculations.

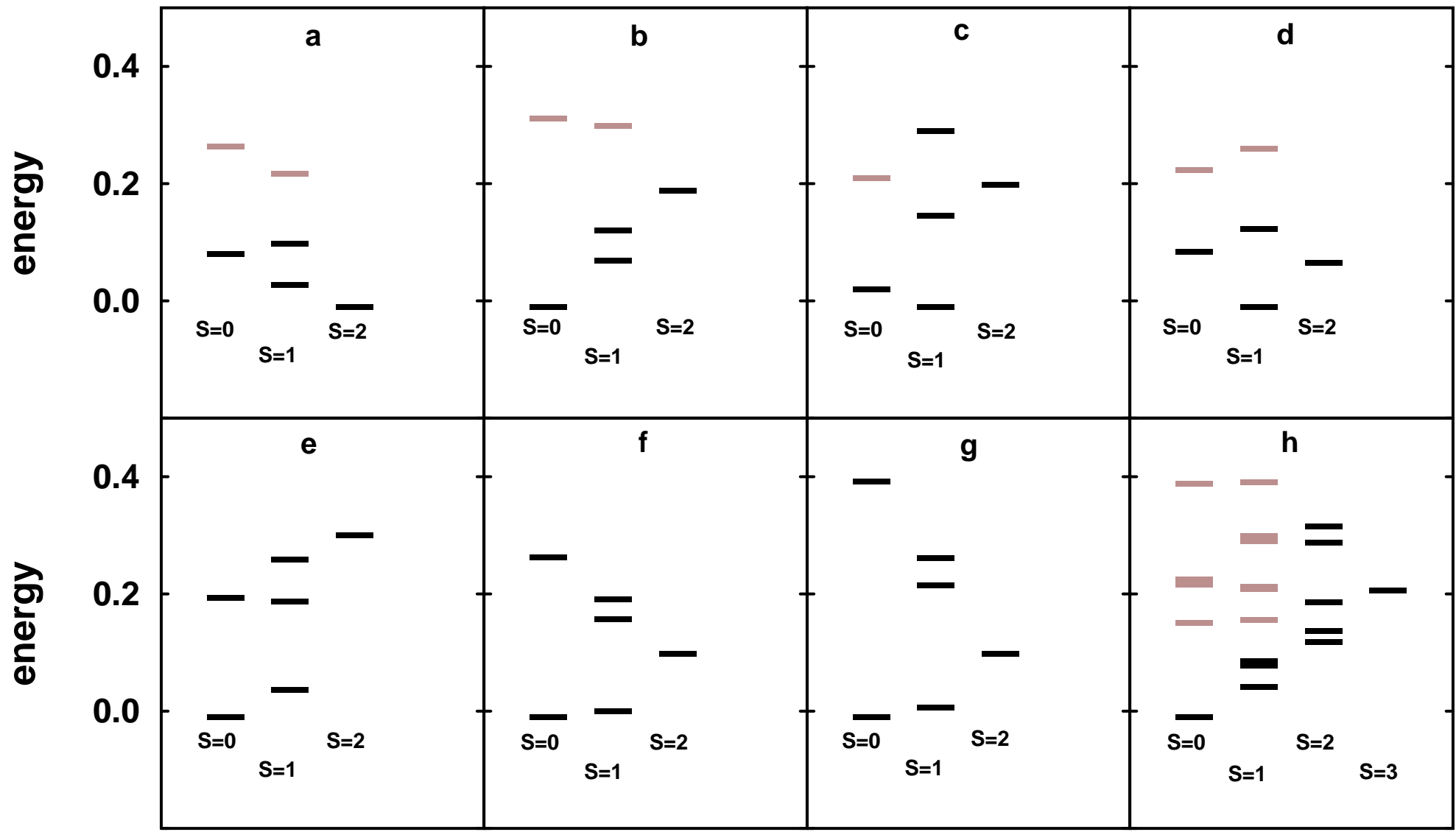
\begin{figure}
\caption{Low energy spectra of the ladder with four and six impurities 
        from DMRG (black lines) and from the effective model (light gray
        lines, only when DMRG data are not available) 
        for different configurations characterized by 
        the sign (F, AF) and the number of complete rungs between the 
        unpaired spins as indicated in the graphs. (a) - (g) four 
        unpaired spins: (a) F4 - F3 - F5, (b) AF4 - F3 - AF5, 
        (c) F4 - AF3 - AF5, (d) F4 - F3 - AF5, (e) AF4 - AF3 - AF5,
        (f) F5 - AF4 - F5, (g) F3 - AF4 - F3;
        (h) six unpaired spins: AF5 - F4 - AF6 - F4 - AF5.} 
        \label{ffimps} 
\end{figure}

In fig.~\ref{ffimps} and table~\ref{dmrgspec} we present a number of
spectra with levels classified according to total spin $S$ and
corresponding results from the effective pair model. We have chosen
the parameters under the following aspects:

\begin{enumerate}
\item We want to control the effective model,

\item we want to illustrate the filling of the gap with defect states, and

\item we want to present the effects of the different sign
combinations for defects on a given sequence of rungs.
\end{enumerate}

For the comparison between the complete ladder spectra as obtained
from the DMRG and the spectra of the effective model we refer to
table~\ref{dmrgspec}. We see that the agreement is very
satisfying, deviations are generally below 10\% with a tendency of
better accuracy for low energies. We note that all DMRG energy
spectra overestimate the true energies. Further corrections will
therefore rather improve the agreement between effective model and
DMRG calculations.

Since the spectra obtained by the two approaches agree so well, in
fig.~\ref{ffimps} DMRG spectra (full lines) only are given when these
are available. However, the higher singlets and triplets were not
accessible at reasonable computational expense as the respective
states are already quite high-lying in the $S^z_{tot} = 0, 1$
sectors. It is only in these cases that we have included into
fig.~\ref{ffimps} the spectra obtained in the effective model (light
gray lines).

In fig.~\ref{ffimps}(a-e) we show spectra for four unpaired spins with
4, 3 and 5 complete rungs in between. The possible combinations of the
signs are seen to lead to quite different spectra and the general
tendency to fill the regular ladder gap is evident.
Figs.~\ref{ffimps}(a, e) result from each other qualitatively by an
overall sign change. Although the magnitudes of the interactions are
somewhat different (compare table~\ref{tabJopen}) this is clearly
evident in these spectra. We therefore present only three out of the
remaining six spectra for this combination of distances in
fig.~\ref{ffimps}(c - e). Fig.~\ref{ffimps}(f, g) presents two
examples of symmetric configurations of four spins (here levels are
characterized by the additional quantum number parity which we have
not indicated). The six-defect spectrum shown in fig.~\ref{ffimps}(h)
again is for a symmetric configuration. In particular these data
illustrate, beyond related calculations, which have been done for the
case of two impurities \cite{MotKFI96}, how the density of states with
$S=1$ may increase at low energies with the number of impurities.
However, a much larger number of impurities will be required to obtain
a reliable numerical estimate for the low energy density of states
$\rho(\epsilon)$.
   
\section{Conclusions} 

We have demonstrated that interacting impurities drastically change
the low-energy spectrum of an antiferromagnetic spin ladder.  A
statistical distribution of impurities reduces the ladder, as far as
its low energy spectrum is concerned (energy range of the pure ladder
gap), to a $S={1 \over 2}$ chain with random interactions, a model
which has recently been used to discuss the expected low temperature
properties of the impure ladder \cite{SigF96}. From our calculations
precise information on the parameters of the effective interactions is
available. For large distances between defects the effective coupling
strength is found to decay purely exponentially. We have found that it
is sufficient to use two spin interactions to describe the spectrum of
the effectively random chain to within 10\%. The spectra which we have
calculated also illustrate the beginning of the process that a random
distribution of impurities with concentration $c$ produces a large
number (of the order of $2^{cL}$) of low-lying states which will fill
the ladder gap as observed experimentally \cite{AzuFTIONT96}.
Applying the Lieb-Mattis theorem to the impure ladder with
antiferromagnetic interactions only we argue that for $T \to 0$ the
susceptibility is characterized by a $1/T$ divergence with finite
Curie constant. Our numerical results, however, are not sufficiently
accurate to draw conclusions about the change in Curie constant with
increasing temperature.

From our results it becomes also clear that interactions between
unpaired spins which are not nearest neighbors and $m-$spin
interactions are present and cannot be neglected in the effective
model at the outset. They have, however, turned out not quantitatively
relevant for the low-energy spectra which we have computed.

The effects of impurities on the spin distribution which we have
presented show some similarity to what has been discussed in the
context of impure Haldane chains \cite{Kat92} and it is useful to
qualitatively compare these two phenomena once more: In both systems
the unpaired spin at the position of the impurity shares its magnetic
moment with the adjacent part of the system - and the degree of this
mixing, i.e.\ its spatial extent, is determined by the correlation
length. In the Haldane case, one deals with a physically different
magnetic impurity spin; it interacts with that part of the unpaired
boundary spin which remains localized, leading to the experimentally
observed splitting of the ESR spectra. In the spin ladder a
nonmagnetic impurity is present and the corresponding unpaired spin
shares its magnetic moment with the adjacent parts of the ladder; the
tails of this magnetic moment introduce an effective interaction
leading to the experimentally observable splitting of the low-energy
spectrum.

\section*{Acknowledgements} 

This work was supported by the German Federal Ministry of Research and
Technology (BMBF) under contract number 03-MI4HAN-8. We wish to thank
Regionales Rechenzentrum Niedersachsen and Zuse Rechenzentrum Berlin
for their helpful cooperation with part of the numerical calculations.

\begin{table}
\caption{Energies calculated by the DMRG before (first line) and after
         one (second line) and two (third line) applications of the 
         finite size algorithm.}
\label{tabdmrg}
\end{table}

\begin{table}
\caption{Effective exchange energies for two unpaired spins at the edge
         and in the bulk of a ladder with $p$ rungs in between. Two energies
         are given for each $p$, corresponding to the two positions on 
         each rung.} 
\label{tabJopen}
\end{table}

\begin{table}
\caption{Low energy spectra of the ladder with four and six impurities 
        for different configurations as in fig.\ref{ffimps}: comparison 
        of DMRG and effective Hamiltonian results.}
\label{dmrgspec}
\end{table}

\end{document}